\documentclass[12 pt,draftcls]{IEEEtran} \onecolumn

\ifCLASSINFOpdf
\else
\fi
\usepackage[dvips]{graphicx}
\usepackage[cmex10]{amsmath}
\usepackage{amssymb}
\usepackage{subfigure}
\usepackage{algorithmic}
\usepackage{array}
\usepackage{mdwmath}
\usepackage{eqparbox}
\usepackage[font=footnotesize]{subfig}
\usepackage{graphicx}
\usepackage{algorithmic}
\usepackage{algorithm}
\usepackage{color}
\usepackage{float}
\usepackage{cite}

\newcommand{\ccc}{\mathbf{c}}
\newcommand{\R}{\mathbf{R}}
\newcommand{\yp}{{\cal P}}
\newcommand{\sss}{\mathbf{s}}
\newcommand{\rrr}{\mathbf{r}}

\newcommand{\qed}{\nobreak \ifvmode \relax \else
      \ifdim\lastskip<1.5em \hskip-\lastskip
      \hskip1.5em plus0em minus0.5em \fi \nobreak
      \vrule height0.75em width0.5em depth0.25em\fi}

\newtheorem{theorem}{Theorem}
\newtheorem{proposition}[theorem]{Proposition}

\begin{document}
%
\title{Permutation Trellis Coded Multi-level FSK Signaling to Mitigate Primary User Interference in Cognitive Radio Networks}

\author{Raghed~El-Bardan,~\IEEEmembership{Student Member,~IEEE,}
        Engin~Masazade,~\IEEEmembership{Member,~IEEE,}
        Onur~Ozdemir,~\IEEEmembership{Member,~IEEE,}
        Yunghsiang~S.~Han,~\IEEEmembership{Fellow,~IEEE,}
        ~and~Pramod~K.~Varshney,~\IEEEmembership{Fellow,~IEEE}
\thanks{R. El-Bardan and P. K. Varshney are with the Department of Electrical Engineering and Computer Science, Syracuse University, Syracuse, NY 13244 USA.}
\thanks{E. Masazade is with the Department of Electrical and Electronics Engineering, Yeditepe University, Turkey.}
\thanks{O. Ozdemir is with Boston Fusion Corp., 1 Van de Graaff Drive, Suite 107, Burlington, MA 01803 USA.}
\thanks{Y. S. Han is with the Department of Electrical Engineering at National Taiwan University of Science and Technology, Taipei, Taiwan. He visited Syracuse University during 2012-13.}
\thanks{Email: \{raelbard, oozdemir, varshney\}@syr.edu, engin.masazade@yeditepe.edu.tr, yshan@mail.ntpu.edu.tw}}

\maketitle

\begin{abstract}
We employ Permutation Trellis Code (PTC) based multi-level Frequency Shift Keying signaling to mitigate the impact of Primary Users (PUs) on the performance of Secondary Users (SUs) in Cognitive Radio Networks (CRNs). The PUs are assumed to be \textit{dynamic} in that they appear intermittently and stay active for an unknown duration. Our approach is based on the use of PTC combined with multi-level FSK modulation to make the SU transmissions robust against PU interference and noise disturbances and help an SU improve its data rate by increasing its transmission bandwidth while operating at low power and not creating destructive interference for PUs. We evaluate system performance by obtaining an approximation for the actual Bit Error Rate (BER) using properties of the Viterbi decoder and carry out a thorough performance analysis in terms of BER and throughput. The results show that the proposed coded system achieves i) \emph{robustness} by ensuring that SUs have stable throughput in the presence of heavy PU interference and ii) \emph{improved resiliency} of SU links to interference in the presence of multiple \textit{dynamic} PUs.
\end{abstract}


\IEEEpeerreviewmaketitle

\section{Introduction}

\IEEEPARstart{T}oday a large portion of the frequency spectrum assigned by the Federal Communications Commission (FCC) is used intermittently resulting in inefficient usage of the assigned frequency bands. In order to accommodate the increased demand for spectrum in wireless applications, Dynamic Spectrum Access (DSA) has been proposed as the new communication paradigm for wireless systems where the existing spectrum is utilized opportunistically \cite{rf1,rf15}. The radios using this technique are called Cognitive Radios (CRs). In a Cognitive Radio Network (CRN), Primary Users (PUs) are the licensed users who have the right to access their spectrum at any time. When Secondary Users (SUs) have data to transmit, they look for spectrum holes which are the frequency bands not being used by primary users and transmit their data over them \cite{rf17,rf16}. In order to ensure a non-degraded PU operation in the network, existing literature has considered several effective protection techniques, e.g., beamforming of SU signals and putting interference temperature limits to satisfy PU signal-to-interference plus noise constraint \cite{bbf,it}. In addition, there are scenarios in which SUs can co-exist with the PUs as long as they do not violate the quality of service requirements of the PUs \cite{itm,itm2}.

Orthogonal frequency division multiplexing (OFDM) has been suggested as a multi-carrier communication candidate for CR systems where the available spectrum is divided into sub-carriers each of which carries a low rate data stream \cite{ofdm,rrf4,rrf7,rrf5,rrf6}. A typical approach for a CR using OFDM is to sense the PU activity over the sub-carriers and then adjust its communication parameters accordingly. The goal is to protect PUs as well as intended SU receivers from possible collisions resulting from the use of the same sub-carriers. Continuous spectrum sensing and re-formation of wireless links may result in substantial performance degradation for SUs \cite{wang}. For instance, the throughput of the secondary system is affected by the time spent for channel sensing. When an SU spends more time on spectrum sensing, a smaller number of information bits will be transmitted over a smaller interval of time resulting in reduced system throughput. On the other hand, decreasing sensing duration may result in a larger probability of making incorrect decisions, thereby decreasing the throughput of both SU and PU. In this regard, the authors in \cite{gozde} numerically analyze the trade-offs between throughput and sensing duration in addition to coding blocklength and buffer constraints. In \cite{liang}, the authors study the problem of designing a sensing slot duration to maximize the achievable throughput for the SUs under the constraint that the PUs are sufficiently protected.

Error correction coding (ECC) has been proposed for CRNs as one possible technique to better protect SU links and prevent degradation in their performance. In \cite{rrf4}, it is assumed that an SU transmitter vacates the band once a PU is detected. Due to the sudden appearance of a PU, rateless codes have been considered to compensate for the packet loss in SU data which is transmitted through parallel subchannels. The authors in \cite{rrf7} consider the design of two efficient anti-jamming coding techniques for the recovery of lost transmitted packets via parallel channels, namely rateless and piecewise coding. Similar to the spectrum model defined in \cite{rrf4}, a system's throughput and goodput are analyzed and a performance comparison is carried out between these two coding techniques. For an OFDM-based CRN presented in \cite{rrf5}, SU transmitters and receivers continuously sense the spectrum, exchange information and decide on the available and unavailable portions of the frequency spectrum. Depending on frequency availability, an appropriate Reed-Solomon coding scheme is used to retrieve the bits transmitted over the unavailable portions of the frequency spectrum. The authors in \cite{rrf6} further explore Low-Density-Parity-Check (LDPC) codes in an OFDM scheme where a switching model is considered for dynamic and distributed spectrum allocation. They also analyze the effects of errors during PU detection on channel capacity and system performance. The switch is assumed to be open for each SU detecting a PU. When the switch is open, the channel is modeled as a binary erasure channel (BEC) and the cognitive transmitter continues to transmit its message allowing bits to be erroneous when received. Another major use of the ECC schemes is presented in \cite{rrf1} where the authors study the performance of cooperative relaying in cognitive radio networks using a rateless coding error-control mechanism. They assume that an SU transmitter participates in PU's transmission as a relay instead of vacating the band in order to reduce the channel access time by a PU. Since the use of rateless codes allows an SU receiver to decode data regardless of which packets it has received as long as enough encoded packets are received, these codes are very suitable for cooperative schemes. The authors in \cite{rrf2} propose an end-to-end hybrid ARQ scheme in CRNs consisting of unidirectional opportunistic links to reduce the number of retransmissions with a fixed throughput offset. Their error control approach is based on coded cooperation among paths and amplify-and-forward relaying of packets within a path such that this hybrid ARQ works for CRNs even if some coded data are missing. The authors implement their scheme using convolutional codes combined with BPSK modulation.

In this paper, we investigate the error correction performance of Permutation Trellis Codes (PTCs) \cite{rf11,rf10} combined with multi-level Frequency Shift Keying modulation systems in CRNs. The proposed PTC based framework is quite general and is applicable to many systems beyond CRNs where interference is an issue. For example, the framework has been applied to power line communications~\cite{rf11,rf10} where strong interferers are assumed to be always present as an extreme case. In this paper, we show a much wider applicability of the PTC based framework as an interference mitigation approach by applying it to CRNs where the special challenge is to guarantee reliable communication by SUs in the presence of intermittent PUs that stay active for an unknown duration without the need of detecting them.  These \textit{dynamic} PUs serve as the source of strong interference for SU communications. Another challenge lies in the susceptibility of SU transmissions in a CRN to intentional interference like jamming attacks. In such a case, it is extremely important to guarantee a minimum degree of operability of the network. In other words, additive white Gaussian noise (AWGN) is not the critical limitation here, but rather an unpredictable and/or intermittent interference, e.g., PU transmissions and jamming attacks. Therefore, the emphasis of the proposed PTC based framework is on robustness, rather than on data rate or bandwidth use. To the best of our knowledge, our work is the first to consider the use of PTCs in CRNs. The motivation for this proposed approach is to overcome the tremendous degrading effect of PU interference on an SU transmission and to provide a stable level of reliable information reception regardless of how severe or prolonged a \textit{dynamic} PU activity is and without requiring an accurate detection of it. In our model, an SU transmits its own information using low power concurrently with PU transmissions without the need to relay PU's traffic \cite{rrf1} or to vacate the band \cite{rrf4}. Different from \cite{rrf5}, we assume that no information exchange or spectrum negotiation takes place between the SU transmitter-receiver pair. Our proposed scheme is different from existing work where communication sessions are carried over parallel subchannels \cite{rrf4,rrf7,rrf5,rrf6}. This is due to the fact that our scheme is based on multi-level FSK modulation with PTC using a single subchannel at a time. By using PTC, continuous channel sensing by the SU transmitter-receiver pair is no longer required, since an appropriate PTC can cope with high levels of PU interference on a given SU link. Thus, SUs no longer suffer from the huge overhead created by continuous sensing of the spectrum.

We consider a similar system model as devised in~\cite{rf11,rf10} and carry out a thorough performance analysis in terms of BERs and throughput for dynamically varying interference. In~\cite{rf11,rf10}, the authors were interested in the design of PTCs that mitigate the impact of permanent interferences that exist in power line communications. In their work, they conducted a simulation study and presented results on the performance of different PTCs with respect to the probability of an element of the code matrix being in error in the presence of different noises on one or more channels. They also investigated the choice of distance increasing/conservative/reducing mappings for a given PTC with respect to narrowband interference that \textit{always} exists on one band in the presence of background noise in terms of BER. The authors in \cite{rf11,rf10} provide a BER performance analysis via simulations and not analytically.

The analytical BER analysis of a PTC coded M-FSK system is imperative to determine its link quality. In other words, the BER estimate is a useful tool for cross-layer design. Given that the PU stays active once it starts transmitting, the exact bit error rate (BER) is derived using an exhaustive search \cite{ragg}. 
Hence, the analytical evaluation of BER becomes computationally prohibitive for large codes. Rather than using an exhaustive search, in this paper, we develop an approximation of BER using the properties of the Viterbi decoder. Also, in our previous work \cite{ragg}, a very special case is considered where a PU stays active once it starts transmission. In this paper, we consider a more practical scenario where a 2-state Markov chain is employed to model the \textit{dynamic} PU activity in terms of alternating On-Off periods. It should be noted that the work on \cite{rf11} and \cite{rf10} also considered an extreme case where the authors assumed that the interference was always present and did not consider dynamically varying interference. In the more general framework presented in this paper, the performance results provided, e.g., BER, are intuitive and more useful in comparison to the worst-case guaranteed performance presented in \cite{ragg}.

Under the same resources such as energy per bit, transmission bandwidth, and transmission time, we compare the throughput performance of the proposed communication system with a simple uncoded opportunistic M-FSK system and a convolutionally coded OFDM system using BPSK modulation. The simulation results show that the proposed system outperforms the aforementioned systems in the presence of heavy PU interference. We also compare the BER performance of the proposed communication system with an LDPC code coupled with an $M$-FSK modulation system for static and dynamic activities of a PU. It is shown that the proposed scheme outperforms the other one. Moreover, we present results on the degree of resiliency to interference that a PTC can guarantee for the SU link in the presence of interference by multiple dynamic PUs. 

To summarize, the main contributions of the paper are:
\begin{itemize}
\item We consider a more practical scenario in which we represent the cognitive radio network using a dynamic PU channel occupancy model including the case of multiple PUs.
\item To mitigate interference created by these dynamic PUs, we propose the use of PTC-based multi-level FSK signaling that incurs no overhead costs as no channel sensing is required before SU transmissions.
\item Using the proposed PTC based framework, the SU transmissions achieves \emph{robustness} against the dynamic PUs' interference and noise disturbances and, at the same time, the \emph{resiliency} of SU links improves.
\item A thorough analytical performance analysis of the proposed scheme is presented in terms of BER and throughput for static and dynamic activities of multiple PUs.
\end{itemize}

The rest of the paper is organized as follows. In Section~\ref{sec:model}, we provide the system model and introduce the concept of PTC. In Section~\ref{sec:analysis}, we introduce hard-decision Viterbi decoding of the system. We also approximate the actual BER through the use of a specified finite number of paths forming the Viterbi trellis. Section~\ref{sec:NRD} presents numerical results demonstrating the improvement in the performance of the PTC-based multi-level FSK system, validating the suggested BER approximation, supporting the effectiveness of the proposed communication scheme, as well as demonstrating the efficiency of PTC in the presence of multiple \textit{dynamic} PU interference. Finally, we conclude the paper and address some future research directions in Section~\ref{sec:conc}.

\section{System Model}
\label{sec:model}

\begin{figure}[t]
\centering
\includegraphics[width=0.3\textwidth,height=!]{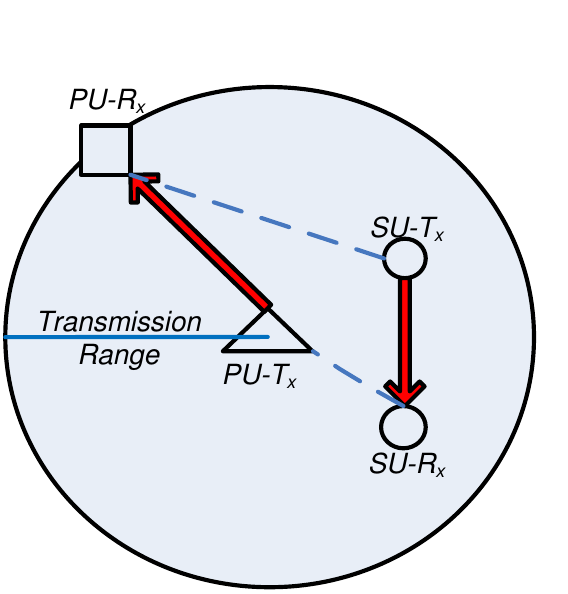}
\caption{\small A simple cognitive radio network.}
\label{fig:fig1}
\end{figure}

\begin{figure}[t]
\centering
\includegraphics[width=1\textwidth,height=!]{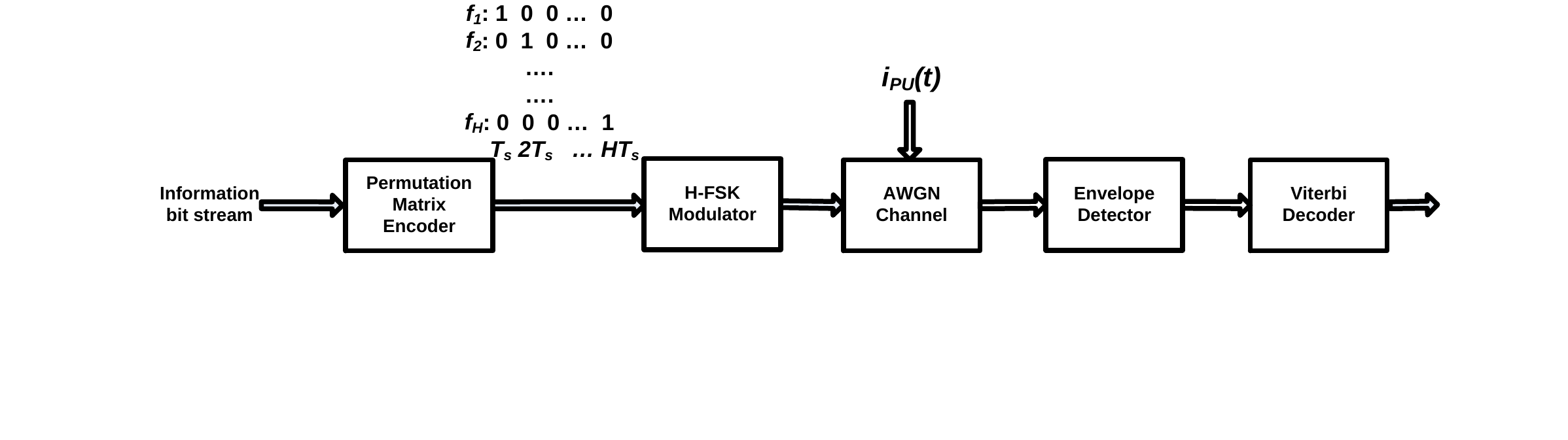}
\caption{\small Block diagram of the Coded multi-level FSK System
}\label{fig:vit}
\end{figure}

We consider a simple cognitive radio network as shown in Fig. \ref{fig:fig1}, which consists of a PU transmitter-receiver pair and an SU transmitter-receiver pair. The SU pair is assumed to be located within the transmission range of the PU.
The transmission range shown in Fig. \ref{fig:fig1} is the maximum distance covered by a PU transmission such that the signal to interference plus noise ratio (SINR) at the PU receiver equals a minimum threshold value, $SINR_{PU}^{*}$. In this model, we assume a free space path loss model and a Line-Of-Sight (LOS) AWGN channel. The power in the transmitted signal is $P_{t}$, so the received power is given in \cite{goldsmith} by
\begin{eqnarray}
\label{pow}
 P_R &=& P_T \cdot \left(\frac{\sqrt{G_l}\lambda}{4\pi d}\right)^2,~
\end{eqnarray}
where $\sqrt{G_{l}}$ is the product of the transmit and receive antenna field gains in the LOS direction and $\lambda$ is the signal wavelength defined as the ratio of the speed of light to the center frequency of the band in operation, $\frac{C}{f_j}$. In this model, we assume omnidirectional antennas so $\sqrt{G_{l}} = 1$. Both licensed and unlicensed users coexist and they can simultaneously operate over the same band. One further assumption made here is that interference, created by an SU, that deteriorates the QoS of the PU is negligible compared to the received PU power. This is due to the fact that the transmission power of PU is much larger than the transmission power of SU, i.e., $P_T^{SU}\ll P_T^{PU}$. In this approach, the SINR at the PU receiver \cite{rfp} given in (\ref{pucond}) satisfies $SINR \ge SINR_{PU}^{*}$, that is,
\begin{eqnarray}
\label{pucond}
SINR &=& \frac{P_R^{PU}}{N_{0} + P_{I}^{SU}} \ge SINR_{PU}^{*},~
\end{eqnarray}
where $P^{SU}_{I}$ is the SU interference at the PU receiver and $SINR_{PU}^{*}$ is the minimum threshold above which a PU transmission is received successfully.
It is assumed that the SU transmitter knows the location of its corresponding receiver possibly by means of extra signaling. An overview of the signal processing model that combines the PTC scheme with the multi-level FSK communication system is provided in Fig. \ref{fig:vit}.
The information bits, $m$, are loaded into a rate $R = \frac{m}{n}$ convolutional encoder in parallel, e.g., $R = \frac{1}{2}$ with a two-stage shift register and generator 7, 5 (octal)~\cite{rf11}. It is well-known that due to the incorporation of a channel coding scheme in a communication system, the overall transmission energy per bit is less than that without a coding scheme. However, due to its error correction capabilities, the purpose of convolutional encoding is to improve the performance of the system. 

For each of the convolutional coded output symbols, the PTC encoder assigns a certain code matrix which is transmitted over both time and frequency. We present an example for illustration, in which the coded symbol ``01'' is mapped onto ``213''. This permutation code matrix is to be transmitted in both time and frequency domains resulting in a $3$ $\times$ $3$ ($H$ $\times$ $H$) binary code matrix as given below.
 \begin{equation}
 \label{ww}
\mathbf{T}_{i} =  \left[
              \begin{array}{cccc}
                0 & 1 & 0 \\
                1 & 0 & 0 \\
                0 & 0 & 1 \\
              \end{array}
            \right],
\end{equation} where $1\le i\le M$. $M = 2^m$ denotes the number of symbols and $H$ defines the number of frequency bands as well as the number of time steps used in transmitting the outputs of the encoder. In this paper, we define the PTC-based multi-level FSK system as the $H$-FSK system since the PTC-based multi-level FSK modulated system leads to an $H$ x $H$ binary code matrix. According to (\ref{ww}), transmission takes place on $f_{2}$, $f_1$, and $f_3$ (first, second and third columns of $\mathbf{T}_i$) corresponding to the time steps $T_s$, $2T_s$, and $3T_s$ respectively. Tables~\ref{table:codeperm} and~\ref{table:code perm} present the symbol (with 1 bit and 2 bits) mappings onto the corresponding unique permutation code matrices respectively~\cite{rf11}. For mapping tables for larger values of $H$, we refer the reader to~\cite{rf11}.
\begin{table}[h]
\caption{Mapping of symbols into permutation code matrices ($\mathit{M} =\mathit{H}$ = 2)}
    \label{table:codeperm}
\begin{center}
    \begin{tabular}{| c | c | c |}
    \hline
Label & Symbol & Permutation code matrix \\ \hline
$\mathbf{T}_1$ & $0$ & $12$\\ \hline
$\mathbf{T}_2$ & $1$ & $21$\\ \hline
\end{tabular}
\end{center}
\end{table}
\begin{table}[h]
\caption{Mapping of symbols into permutation code matrices ($\mathit{M}$ = 4, $\mathit{H}$ = 3)}
    \label{table:code perm}
\begin{center}
    \begin{tabular}{| c | c | c |}
    \hline
Label & Symbol & Permutation code matrix \\ \hline
$\mathbf{T}_1$ & $00$ & $231$\\ \hline
$\mathbf{T}_2$ & $01$ & $213$\\ \hline
$\mathbf{T}_3$ & $10$ & $132$ \\ \hline
$\mathbf{T}_4$ & $11$ & $123$ \\ \hline
\end{tabular}
\end{center}
\end{table}
For $M$ different symbols, $\mathbf{T}_i$ denotes the set of transmitted code matrices such that  $\mathbf{T}_i \in$ $\{\mathbf{T}_{1},\ldots,\mathbf{T}_{M}\}$ has the following general form:
  \begin{equation}
  \label{T-i}
\mathbf{T}_i =  \left[
              \begin{array}{cccc}
                q_{1,1} & q_{1,2} & \ldots & q_{1,H} \\
                : & \ddots & \ldots & : \\
                q_{H,1} & \ldots & \ldots & q_{H,H} \\
              \end{array}
            \right],~
\end{equation}
where $q_{j,k} \in \{0, 1\}$ denotes the ($j, k$) binary element in the $i$-th transmitted code matrix, $j$ indicates the output of the detector for frequency $f_j$ at time step $k$ in the code matrix. At each time step, the element where 1 is present in the code matrix is modulated and transmitted. At a given time, only one frequency band is used for transmission.
Using the $H$-FSK scheme, the transmitted signal, assumed to be sufficiently narrowband, over $j$-th frequency and $k$-th time step where $j,k  \in \{1,2,\ldots,H\}$, is given by:
\begin{eqnarray}
\label{signal_eq}
s_{j}(t) &=& \sqrt{{2\frac{E_{s}/H}{T_{s}}}}\cdot\cos(2{\pi}f_{j}t), \quad (k-1)T_s\le t\le k T_{s}, \nonumber \\
f_{j} &=& f_{1} + \frac{j - 1}{T_s}, \quad 1\le j\le H,
\end{eqnarray}
where $E_{s} = P_{T}^{SU} \cdot T_s$ is the transmitted signal energy per information symbol and $T_{s}$ is the symbol duration. In this model, we assume a strong LOS path between a stationary SU transmitter and a stationary receiver. Therefore, channel noise is modeled as Additive White Gaussian Noise (AWGN) with zero-mean and variance $\frac{N_{0}}{2}$. Since path loss is a function of the operating frequency, we need to ensure that the received signal power over the different frequency channels is the same, $P_{R}^{SU}(f_{1}) = \ldots = P_{R}^{SU}(f_{H}) = P_{R}^{SU}$, in order to satisfy a fixed signal-to-noise ratio, $E_s / N_0$ at the receiver. Therefore, we adjust each SU's transmitting power over each frequency, $P_{T}^{SU}(f_{j})$, according to the free space path loss model given in (\ref{pow}). Therefore, the SU transmitter needs to know the location of its corresponding receiver. At the input of the SU demodulator, the received signal is given by:
$$x_{j}(t) = \left\{ \begin{array}{lr}
 s_{j}^{r}(t) + i_{PU}^{r}(t) + w(t), &\mbox{if }\exists \mbox{ PU on } f_j, \\
 s_{j}^{r}(t) + w(t), &\mbox{ otherwise,}
       \end{array} \right.$$
where $s_{j}^{r}(t) = \sqrt{2 \frac{E_{s}^{r}/H}{T_s}}\cos (2\pi f_{j}t + \theta)$, $i_{PU}^{r}(t) = \sqrt{2 \frac{I_{PU}/H}{T_s}}\cos (2\pi f_{j}t + \phi)$, and $E_s^r \triangleq P_{R}^{SU} T_s$ is defined as the symbol energy at the receiver. $I_{PU} \triangleq P_I^{PU} T_s$ and $P_I^{PU}$ are defined as the interference energy per coded symbol and the interference power due to the PU transmitter at the SU receiver, respectively, and $w(t)$ represents the channel noise at the receiver. It should be mentioned that a PU transmitter can modulate the transmitted signal using any scheme and is not limited to the use of FSK-modulated signals. The fact that the PU power is very high compared to that of the SU implies that the received signal is dominated by the PU signal which, in turn, shapes the behavior of the system. Without loss of generality, we assume that the PU uses an FSK-modulated signal to transmit its information similar to the SU.

At the receiver side, non-coherent detection is employed using a bank of $H$ quadrature receivers so that each consists of two correlation receivers corresponding to the in-phase and quadrature components of the signal. The in-phase component of the signal received, $x_{I_{j,k}}$, is given by:
\begin{eqnarray}
\label{nnn}
&& x_{I_{j,k}} = \nonumber \\
&& \left\{ \begin{array}{cr}
 \frac{E_s^r}{H} \cos \theta + \frac{I_{PU}}{H}\cos \phi + w, &\mbox{if PU exists on $f_j$ at time $k$,} \\
 \frac{E_s^r}{H} \cos \theta + w, &\mbox{otherwise,}
       \end{array} \right.
\end{eqnarray}
where $\theta$ and $\phi$ are uniformly distributed over [0, 2$\pi$], i.e., $\theta$, $\phi \sim \mathcal{U}(0, 2\pi)$, and denote the random phase components of the SU and the PU signals, respectively. The noise term $w$ in (\ref{nnn}) is modeled as AWGN, i.e., $w \sim \mathcal{N}(0,\frac{N_0}{2})$. $E_s^r$ is the received symbol energy over a given signaling interval. Similarly, $x_{Q_{j,k}}$ is the received signal's quadrature component defined as follows:
\begin{eqnarray}
\label{nnnn}
&& x_{Q_{j,k}} = \nonumber \\
&& \left\{ \begin{array}{cr}
 \frac{E_s^r}{H} \sin \theta + \frac{I_{PU}}{H}\sin \phi + w, &\mbox{if PU exists on $f_{j}$ at time $k$,} \\
 \frac{E_s^r}{H} \sin \theta + w, &\textrm{otherwise.}
       \end{array} \right.
\end{eqnarray}
The envelope of each quadrature receiver over frequency $j$ and time step $k$, $l_{j,k}$, is defined as the square root of the sum of the squared in-phase and quadrature components of the correlator output as
  \begin{eqnarray}
\label{lk}
l_{j,k} = \sqrt{x_{I_{j,k}}^2 + x_{Q_{j,k}}^2}.
\end{eqnarray}
At the receiver, a hard decision decoding scheme is used where the envelope value of each of the $H$ quadrature receivers is compared to a threshold value, $l_{th}$. The threshold value we use in this paper is the same as that used by the authors in \cite{rf11}, namely $l_{th} = 0.6\sqrt{E_s^r}$.

The received code matrix $\mathbf{R}_{i}$ is of the form,
 \begin{equation}
\label{R-n}
\mathbf{R}_{i} =  \left[
              \begin{array}{cccc}
                b_{1,1} & b_{1,2} & \ldots & b_{1,H} \\
                : & \ddots & \ldots & : \\
                b_{H,1} & \ldots & \ldots & b_{H,H} \\
              \end{array}
            \right],~
\end{equation} where $b_{j,k}$ $\in \{0,1\}$ and can be determined from,
\begin{eqnarray}
b_{j,k} &=&
\left\{ \begin{array}{cr}
 1, &\mbox{$l_{j,k}$ $\ge$ $l_{th}$,} \\
 0, &\textrm{otherwise.}
       \end{array} \right.
\end{eqnarray}


\section{Bit Error Rate Analysis}
\label{sec:analysis}
The link quality between two SUs is determined by how much PU interference it can tolerate. We employ BER as the QoS metric to characterize successful communication for an SU.  We consider the system model presented in Fig.~\ref{fig:vit} and obtain an approximation for BER. This system model is the same as the one considered in~\cite{rf11,rf10} where the PTC encoder includes a convolutional encoder and the corresponding decoder uses Viterbi algorithm for decoding.

Here, we assume that the PTC encoder uses a rate $\frac{1}{2}$ convolutional code whose output is converted to a symbol which, in turn, is mapped onto a permutation matrix to be transmitted. It is important to point out that the exact BER analysis of the proposed system is computationally prohibitive. 
Hence, we approximate BER using some properties of the Viterbi decoder.

\begin{figure}[t]
\centering
\includegraphics[width=0.3\textwidth,height=!]{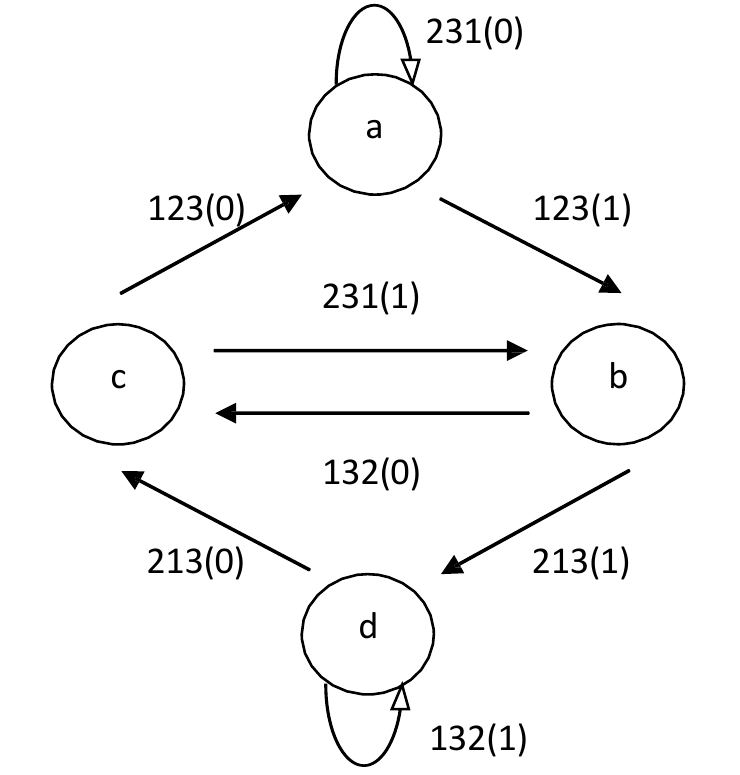}
\caption{\small Finite State Machine of a ($R$ = $\frac{1}{2}$) convolutional coded PTC system}
\label{fig:fig11}
\end{figure}

\begin{figure}[t]
\centering
\includegraphics[width=0.7\textwidth,height=!]{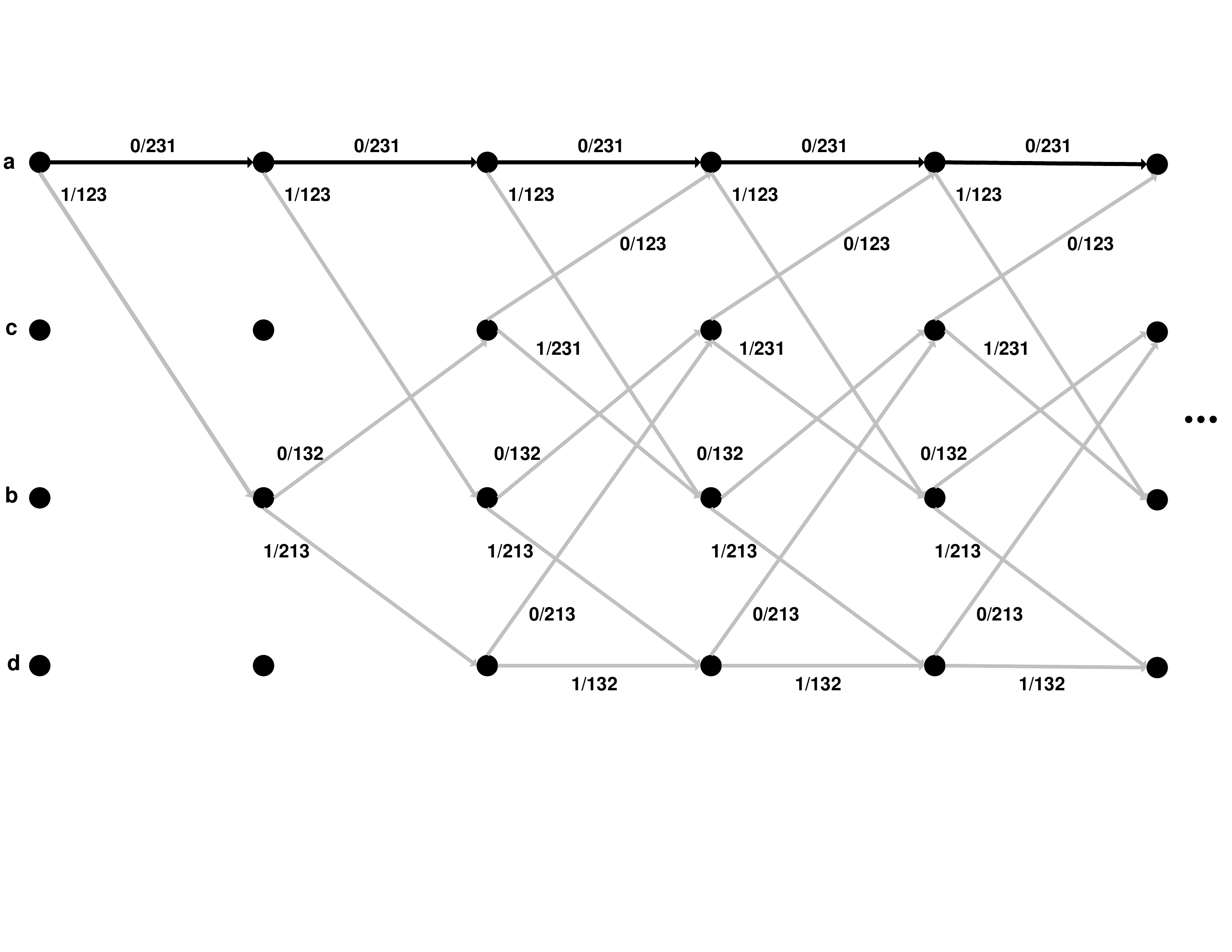} \vspace{-2cm}
\caption{\small A ($R$ = $\frac{1}{2}$) convolutional code trellis for our PTC-based system.}
\label{fig:trr}
\end{figure}

The encoder of the binary convolutional code with a given rate is viewed as a finite state machine as shown in Fig. \ref{fig:fig11}.
  This state diagram, corresponding to the PTC shown in Table \ref{table:code perm}, results in the trellis presented in Fig. \ref{fig:trr}.
In order to optimally decode the received information, we employ the Viterbi algorithm which reconstructs the maximum-likelihood path given the input information sequence. The Viterbi decoder gives as an output the path in the trellis that has the minimum overall Hamming distance with respect to the received sequence. In this regard, an error event may occur if for a transmitted path in the trellis the overall number of differences with the demodulator outputs is larger than or equal to that of a competing path. 

In order to compute an upper-bound on the probability of bit error, all error events have to be taken into account. It is important to note that the number of error events is proportional to the number of input bits to be transmitted. In other words, the number of competing paths and the number of error events increase as more input bits are serially loaded into the convolutional encoder. Then, the probability of bit error of a convolutional code is upper-bounded according to \cite{rf20,LIN83} as
\begin{equation}
\label{berrr}
P_{e} \le \sum_{d = d_{free}}^{\infty} a_dP_2(d),~
\end{equation} where $d_{free}$ is the free distance of the convolutional code, $a_d$ is the number of paths that differ by $d$ bits from the transmitter codeword, and $P_2(d)$ is the probability that the decoded path differs by $d$ bits from the transmitted codeword. In particular, $a_d$ and $P_2(d)$ are independent of the transmitted codeword. In fact, they can be calculated assuming that an all-zero codeword is transmitted. Knowing that all the code matrices to be sent include $H$ non-zero elements, we next prove that there is a similar upper-bound on the system given in Fig.~\ref{fig:vit}.

Now, let $\ccc$ be a codeword of the convolutional code and $\yp(\ccc)$ be the corresponding expanded codeword comprising the permutation code matrices mapped from $\ccc$. The length of $\yp(\ccc)$ is $(L+m)H^2$, where $L$ denotes the size of packet to be transmitted and $L+m$ the number of branches in the trellis. In order to apply (\ref{berrr}), we need to prove that the upper-bound given in~\eqref{berrr} is independent of the transmitted expanded codeword. Let $P_e(\yp(\ccc))$ denote the probability of bit error when $\yp(\ccc)$ is the transmitted codeword. Knowing that $\yp(\ccc)$ is a one-to-one mapping from the codeword $\ccc$, by the argument of \eqref{berrr} given in~\cite{rf20,LIN83}, we have
\begin{eqnarray}
\label{berr2}
P_e(\yp(\ccc))&&\le  \sum_{d = d^*_{free}}^{\infty} a_d^{\yp(\ccc)}P^{\yp(\ccc)}_2(d),~
\end{eqnarray}
where $d^*_{free}$ is the free distance of the expanded code, and $a_d^{\yp(\ccc)}$ and $P^{\yp(\ccc)}_2(d)$ are the number of paths and the probability that the paths differ by $d$ bits from the transmitted codeword $\yp(\ccc)$, respectively.
\begin{proposition}\label{prop}
The bound in \eqref{berr2} is independent of the transmitted codeword if the noise caused by the PU in a subchannel will affect all bits in this channel that are generated by one or more trellis branches.
\end{proposition}
\begin{IEEEproof}
See proof in Appendix.
\end{IEEEproof}

The fact that we have no information regarding the length of the transmitted bits sequence means that the calculation of the upper bound, in (\ref{berrr}), involves an infinite sum over all possible error events. In other words, an infinite number of trellis' paths have to be taken into consideration which will render the process of evaluating the error bound computationally prohibitive for applications such as link adaptation. Therefore, it is important to approximate BER using a suitably selected finite number of trellis' paths.

For every convolutional code, there exists a transfer function that uniquely determines the number of paths by which a received sequence is $d$ bits different from an assumed all-zero transmitted bit sequence. For instance, the transfer function of the mapped permutation code in the proposed $H$-FSK scheme based on Fig. \ref{fig:fig11}, i.e. $H$ = 3, is given by $\mathit{T(D)}$ where
\begin{equation}
T(D) = D^{16} + 2\cdot D^{20} + 4\cdot D^{24} + 8\cdot D^{28} + \cdots
\end{equation}
Equivalently, we are saying that there is only one path, $\{a \rightarrow b, b \rightarrow c, c \rightarrow a\}$, with an input sequence $\{1,0,0\}$ and permutation code matrix sequence $\{123,132,123\}$ that has 16 bits different from the assumed all-zero transmitted bit sequence ($d_H$ (231,123) + $d_H$ (231,132) + $d_H$ (231,123) = 6 + 4 + 6), 2 paths that have 20 bits difference, and so on. It is most likely that the first few error events (paths) in the trellis dominate all the remaining ones. In this manner, the approximate BER can be written as
\begin{eqnarray}
\label{apxpe}
\hat{P_{e}} && \approx \sum_{d = d^*_{free}}^{d_{free}^*+{z}} a_d^{\yp(\ccc)}P^{\yp(\ccc)}_2(d ),~
\end{eqnarray} where $z \in \mathbb{N}$ and $z+1$ denotes the number of paths involved in the BER approximation for which numerical results are presented in Section~\ref{sec:NRD}.

In order to compute $P^{\yp(\ccc)}_2(d )$ we denote by $\mathbf{T}^u$ and $\mathbf{R}^u$ the transmitted and received code matrices at stage $u$ in a trellis composed of $\mathcal{V}$ stages. Then, $P^{\yp(\ccc)}_2(d)$ can be expressed as,
\begin{eqnarray}
\label{p2p}
P^{\yp(\ccc)}_2(d) &=& \prod_{u=1}^{\mathcal{V}} P\left(\mathbf{R}^u|\mathbf{T}^u\right) \nonumber \\
&=& \prod_{u=1}^{\mathcal{V}} \prod_{j=1}^{H} \prod_{k=1}^{H} P\left(b^u_{j,k}|q^u_{j,k}\right) \nonumber \\
&=& \prod_{u=1}^{\mathcal{V}} \prod_{j=1}^{H} \prod_{k=1}^{H} P\left(b^u_{j,k}|q^u_{j,k},PU_j\right)P\left(PU_j\right) + P\left(b^u_{j,k}|q^u_{j,k},\bar{PU}_j\right)P\left(\bar{PU}_j\right),~
\end{eqnarray} where $q^u_{j,k}$ and $b^u_{j,k}$ are the $(j,k)$ binary elements given by (\ref{T-i}) and (\ref{R-n}) respectively in the $u$-th stage of the trellis and $PU_j$ and $\bar{PU}_j$ denote the presence and absence of PU activity over channel $f_j$ respectively. Here we note that $P(PU_j)$ and $P(\bar{PU}_j)$ depend on channel $f_j$'s occupancy model which, in turn, is discussed in Section~\ref{sec:lcdom}.

In each stage of the trellis, the absence and presence of PU need to be taken into consideration when computing the likelihoods in (\ref{p2p}) that also depend on the code matrix to be transmitted.

\subsection{Computation of likelihoods in the absence of PU}
Suppose there is no PU transmission over frequency band $j$ at time step $k$. If the SU is active, i.e. transmitting a bit `1', then the demodulator in-phase component $x_{I_{j,k}} \sim \mathcal{N}(\sqrt{\frac{{E_{s}^r}}{H}}\cos \theta, \frac{N_{0}}{2})$ and the demodulator quadrature component $x_{Q_{j,k}} \sim \mathcal{N}(\sqrt{\frac{{E_{s}^r}}{H}}\sin \theta, \frac{N_{0}}{2})$. The fact that $x_{I_{j,k}}$ and $x_{Q_{j,k}}$ are statistically independent random variables with non-zero means, then implies $l_{{j,k}} \sim Rice(\sqrt{{\frac{{E_{s}^r}}{H}}}, \sqrt{\frac{N_{0}}{2}})$. Accordingly, $P\left(b_{j,k} = 0 | q_{j,k} = 1,\bar{PU}_j\right)$ can be computed as
\begin{eqnarray}
\label{no10}
P\left(b_{j,k} = 0 | q_{j,k} = 1,\bar{PU}_j\right) &=& P(l_{{j,k}} < l_{th} | q_{j,k} = 1, \bar{PU}_j) \nonumber \\
&=& F_{L_{j,k}}(l_{th}) \nonumber \\
&=& 1 - Q_{1}\Big(\sqrt{2\frac{E^r_{s}/H}{N_{0}}},0.6\sqrt{2\frac{E^r_{s}}{N_{0}}}\Big),~
\end{eqnarray}
where $F_{L_{j,k}}(l_{th})$ is the cumulative distribution function of $l_{j,k}$ evaluated at $l_{th}$, $Q_{1}(v,w)$ is the Marcum's Q-function defined as \cite{book_s},
\begin{eqnarray}
\label{mq}
Q_{1}(v,w) = \int^{\infty}_w x\exp\left\{-\frac{x^2 + v^2}{2}\right\}I_{0}(vx)\,dx
\end{eqnarray}
and $I_0(vx)$ is the zeroth order modified Bessel function. $P\left(b_{j,k} = 1 | q_{j,k} = 1,\bar{PU}_j\right)$, in this case, is nothing but the complement of~(\ref{no10}) and is given by
\begin{eqnarray}
P\left(b_{j,k} = 1 | q_{j,k} = 1,\bar{PU}_j\right) &=& P(l_{{j,k}} \ge l_{th} | q_{j,k} = 1, \bar{PU}_j) \nonumber \\
&=& 1 - F_{L_{j,k}}(l_{th}) \nonumber \\
&=& Q_{1}\Big(\sqrt{2\frac{E^r_{s}/H}{N_{0}}},0.6\sqrt{2\frac{E^r_{s}}{N_{0}}}\Big).
\end{eqnarray}
If the SU is not transmitting any information on subchannel $j$ at time step $k$, then, $x_{I_{j,k}} \sim \mathcal{N}(0, \frac{N_{0}}{2})$ and  $x_{Q_{j,k}} \sim \mathcal{N}(0, \frac{N_{0}}{2})$, and we have $l_{{j,k}} \sim Rayleigh(\sqrt{\frac{N_{0}}{2}})$. Then, $P(b_{j,k} = 1 | q_{j,k} = 0,\bar{PU}_j)$ and $P(b_{j,k} = 0 | q_{j,k} = 0,\bar{PU}_j)$ can be computed as,
\begin{eqnarray}
\label{no01}
P(b_{j,k} = 1 | q_{j,k} = 0,\bar{PU}_j) &=& P(l_{{j,k}} \ge l_{th} | q_{j,k} = 0,\bar{PU}_j) \nonumber \\
&=& 1 - F_{L_{j,k}}(l_{th}) \nonumber \\
&=& \exp \Big(-0.36 \frac{E^r_{s}}{N_{0}}\Big)
\end{eqnarray}
and
\begin{eqnarray}
\label{no01}
P(b_{j,k} = 0 | q_{j,k} = 0,\bar{PU}_j) &=& P(l_{{j,k}} < l_{th} | q_{j,k} = 0,\bar{PU}_j) \nonumber \\
&=& F_{L_{j,k}}(l_{th}) \nonumber \\
&=& 1 - \exp \Big(-0.36 \frac{E^r_{s}}{N_{0}}\Big).
\end{eqnarray}

\subsection{Computation of likelihoods in the presence of PU}
In the presence of PU activity and an active SU, i.e., the SU is transmitting a `1', the signal received at the input of the demodulator has two significant terms besides noise. One corresponds to the actual signal energy of the transmitted symbol and the other is created by the PU activity. It is assumed that the SU transmitting power is sufficiently small so that QoS of the PU session is maintained. Therefore, the SU signal is negligible as compared to that of the PU. In this case, $x_{I_{j,k}} \sim \mathcal{N}(\sqrt{\frac{I_{PU}}{H}}\cos \phi, \frac{N_{0}}{2})$ and $x_{Q_{j,k}} \sim \mathcal{N}(\sqrt{\frac{I_{PU}}{H}}\sin \phi, \frac{N_{0}}{2})$. Thus, $l_{j,k} \sim Rice(\sqrt{\frac{I_{PU}}{H}},\sqrt{\frac{N_{0}}{2}})$. $P(b_{j,k} = 0 | q_{j,k} = 1,PU_j)$ and $P(b_{j,k} = 1 | q_{j,k} = 1,PU_j)$, in this scenario, are calculated according to,
\begin{eqnarray}
\label{p10}
P(b_{j,k} = 0 | q_{j,k} = 1,PU_j) &=& P(l_{j,k} < l_{th} | q_{j,k} = 1,PU_j) \nonumber \\
&=&  F_{L_{j,k}}(l_{th}) \nonumber \\
&=& 1 - Q_{1}\Big(\sqrt{2\frac{I_{PU}/H}{N_{0}}},0.6\sqrt{2\frac{E_s^r}{N_{0}}}\Big)
\end{eqnarray}
and
\begin{eqnarray}
P(b_{j,k} = 1 | q_{j,k} = 1,PU_j) &=& P(l_{j,k} \ge l_{th} | q_{j,k} = 1,PU_j) \nonumber \\
&=&  1 - F_{L_{j,k}}(l_{th}) \nonumber \\
&=& Q_{1}\Big(\sqrt{2\frac{I_{PU}/H}{N_{0}}},0.6\sqrt{2\frac{E_s^r}{N_{0}}}\Big).
\end{eqnarray}
If the SU does not transmit on subchannel $j$ at time step $k$, the output of the demodulator has the noisy signal received from the PU. In that case, and similar to the analysis approach followed earlier, $x_{I_{j,k}} \sim \mathcal{N}(\sqrt{\frac{I_{PU}}{H}}\cos \phi, \frac{N_{0}}{2})$ and  $x_{Q_{j,k}} \sim \mathcal{N}(\sqrt{\frac{I_{PU}}{H}}\sin \phi, \frac{N_{0}}{2})$. Consequently, $l_{j,k} \sim Rice(\sqrt{\frac{I_{PU}}{H}},\sqrt{\frac{N_{0}}{2}})$. So, the following probabilities $P(b_{j,k} = 0 | q_{j,k} = 0,PU_j)$ and $P(b_{j,k} = 1 | q_{j,k} = 0,PU_j)$ are the same as $P(b_{j,k} = 0 | q_{j,k} = 1,PU_j)$ and $P(b_{j,k} = 1 | q_{j,k} = 1,PU_j)$, respectively.

Once the probabilities presented above are evaluated, we can calculate $P(\mathbf{R}_{i'}|\mathbf{T}_{i})$ for each stage of the $\mathcal{V}$ stages in the trellis and, eventually, approximate the BER of the proposed scheme as given in~(\ref{apxpe}).

\section{Throughput Analysis}
\label{sec:through}
So far, we have computed the BER approximately and evaluated the performance of the $H$-FSK system in the presence of PU interference. We have also proposed the adoption of the PTC-based multi-level FSK in the transmission strategy of SU in order to provide resilience to spontaneous narrowband PU interference. In other words, heavy redundancy introduced by the PTC ensures successful communication sessions of the SUs in the presence of PU interference. In order to further explore the performance of the proposed $H$-FSK communication system and validate its effectiveness, we perform a throughput analysis based on the approximate BER provided in Section~\ref{sec:analysis}. Accordingly, an SU, adopting the proposed $H$-FSK communication scheme, is expected to provide a level of reliable information reception which is better than what the SU could achieve otherwise under heavy PU interference.
\subsubsection{Licensed Channel Dynamic Occupancy Model}
\label{sec:lcdom}
\begin{figure}[t]
\centering
\includegraphics[width=0.5\textwidth,height=!]{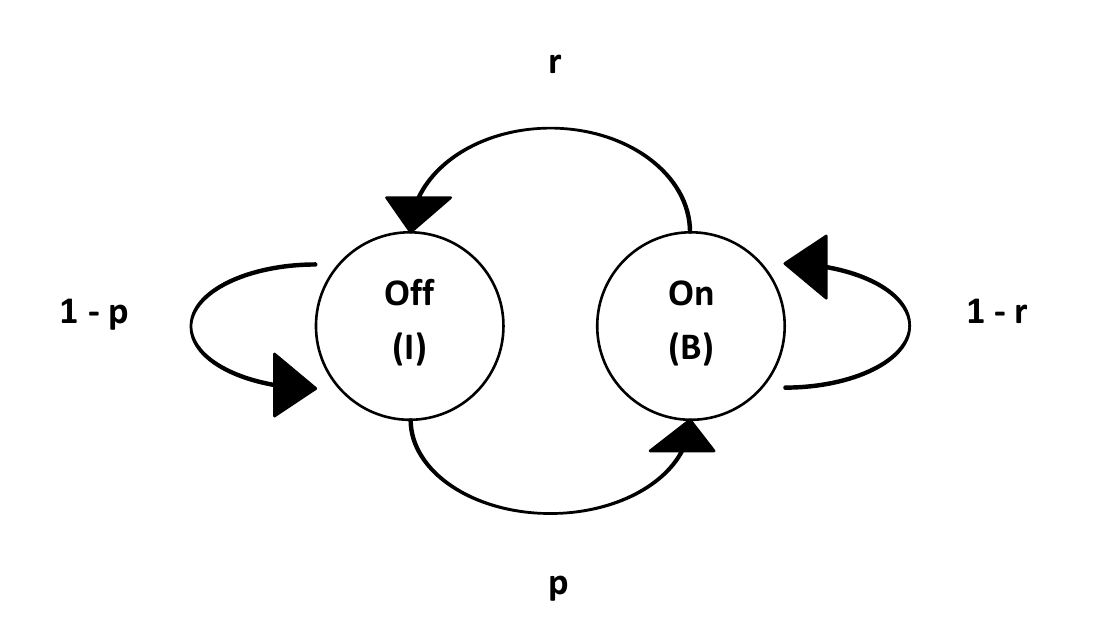}
\caption{\small Two-state Markov chain model for PU activity over licensed spectrum
}\label{fig:marchai}
\end{figure}

Different from \cite{ragg} where the licensed channel was assumed to be \textit{always} occupied by the PU, we model the channel available for PU active transmissions and its occupancy as a 2-state Markov chain as shown in Fig. \ref{fig:marchai}.
  In this model, the channel has two alternative states denoted by On (Busy) and Off (Idle). This assumption is practical in the sense that the licensed spectrum occupancy will experience alternating On-Off periods rather than the assumption of the channel being \textit{always} occupied (PU being \textit{always} present once it starts transmission during an SU transmission). The periods during which the licensed spectrum's state is idle or busy, also known as holding times, are geometrically distributed with known independent parameters $r$ and $p$ as shown in Fig. \ref{fig:marchai}. An On state represents the state in which the licensed band is occupied by a PU resulting in degraded BER performance for SU transmissions, while an Off (idle) state represents the state in which the primary band is idle. Accordingly, the steady state probabilities of the primary channel's state being \emph{Off} or \emph{On} are given respectively as
\begin{equation}
\label{ssof}
{P_{ch}^{Off}} = \frac{r}{r + p}
\end{equation} and
\begin{equation}
\label{sson}
{P_{ch}^{On}} = \frac{p}{r + p}.
\end{equation}
\subsubsection{Throughput Analysis}
\label{sec:through-ana}
In order to examine the effectiveness of the proposed $H$-FSK approach, we conduct a throughput analysis that monitors the average throughput of the SU communication session in the presence of PU activity modeled as a 2-state Markov chain. Knowing that the SU transmitter has no information regarding the presence or absence of the PU, we determine the unlicensed throughput for SU communications as a function of the probability of a licensed channel being occupied by a PU. Such analysis illustrates the efficiency of the proposed scheme in ensuring a stable throughput regardless of any PU activity over the licensed spectrum. Accordingly, we present the approximate expected throughput ($T_{e}$) of the $H$-FSK communication scheme based on the approximate BER given as
\begin{equation}
T_{e} \approx \Big(1 - \hat{PER}\Big)\times R_p \times L, \label{eq:expth}
\end{equation}
where $\hat{PER} = 1-(1-\hat{P_{e}})^L$ is the approximate packet error rate (PER), $\hat{P_{e}}$ is the approximate BER computed with $z+1$ paths knowing the presence of \textit{dynamic} PU interference as described in (\ref{apxpe}) and (\ref{p2p}), and $R_p$ is the rate at which information packets are sent. Throughput, in our model, is defined as the number of correctly received packets per unit time. For simplicity, we do not consider header and overhead bits when computing the link throughput.
\section{Numerical Results and Discussion}
\label{sec:NRD}
\subsection{Approximate BER with Single Permanent PU Interference}
For the coded $H$-FSK communication system shown in Fig. \ref{fig:vit}, the following values of the physical parameters are assumed. The transmit power of the SU is varied between $[25\: {\mu}W, 4\:mW]$ and the noise power density is selected as $N_0 = 2.5\times 10^{-14}$. These values are chosen in such a way so as to not create destructive interference to the PU. We further assume that the lowest frequency band in the available frequency spectrum, $f_1$, is selected as 56 MHz, and the bandwidth spacing between any two subchannels is selected as 6 MHz. The PU is assumed to be operating at $f_2$ and has a total transmitting power of $1$ MW. The parameters' values used in the simulation are chosen in accordance with the IEEE 802.22 standard \cite{ww}. The distances between the PU transmitter and SU receiver and that between the SU transmitter-receiver pair are selected as 10 m, in order to analyze the quality of the SU link under a high interference scenario.

\begin{figure}[t]
\centering
\includegraphics[width=0.6\textwidth,height=!]{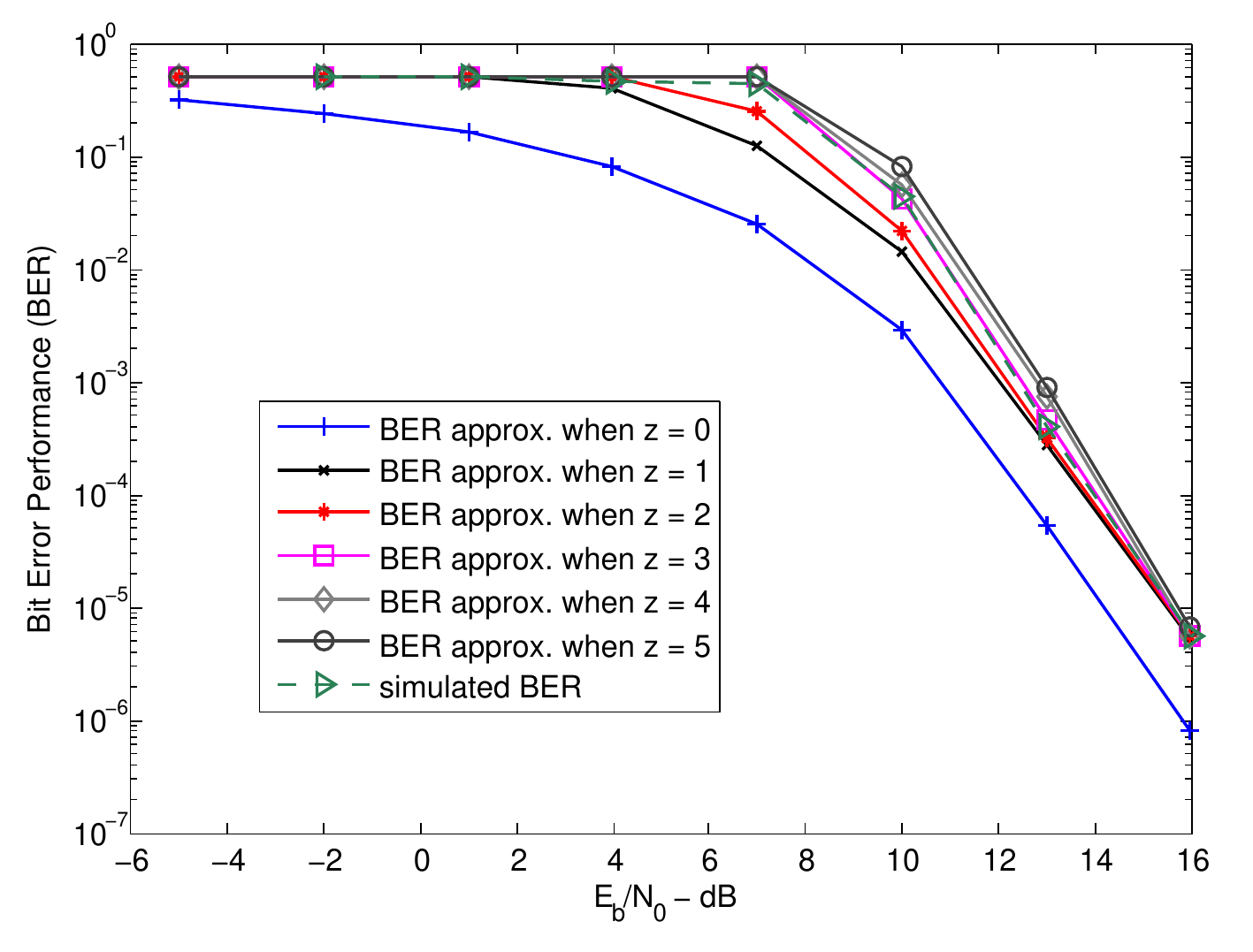}
\caption{\small BER approximations for PTC with $\mathit{H}$ = 3.}\label{fig:upr_b}
\end{figure}

\begin{figure}[t]
\centering
\includegraphics[width=0.6\textwidth,height=!]{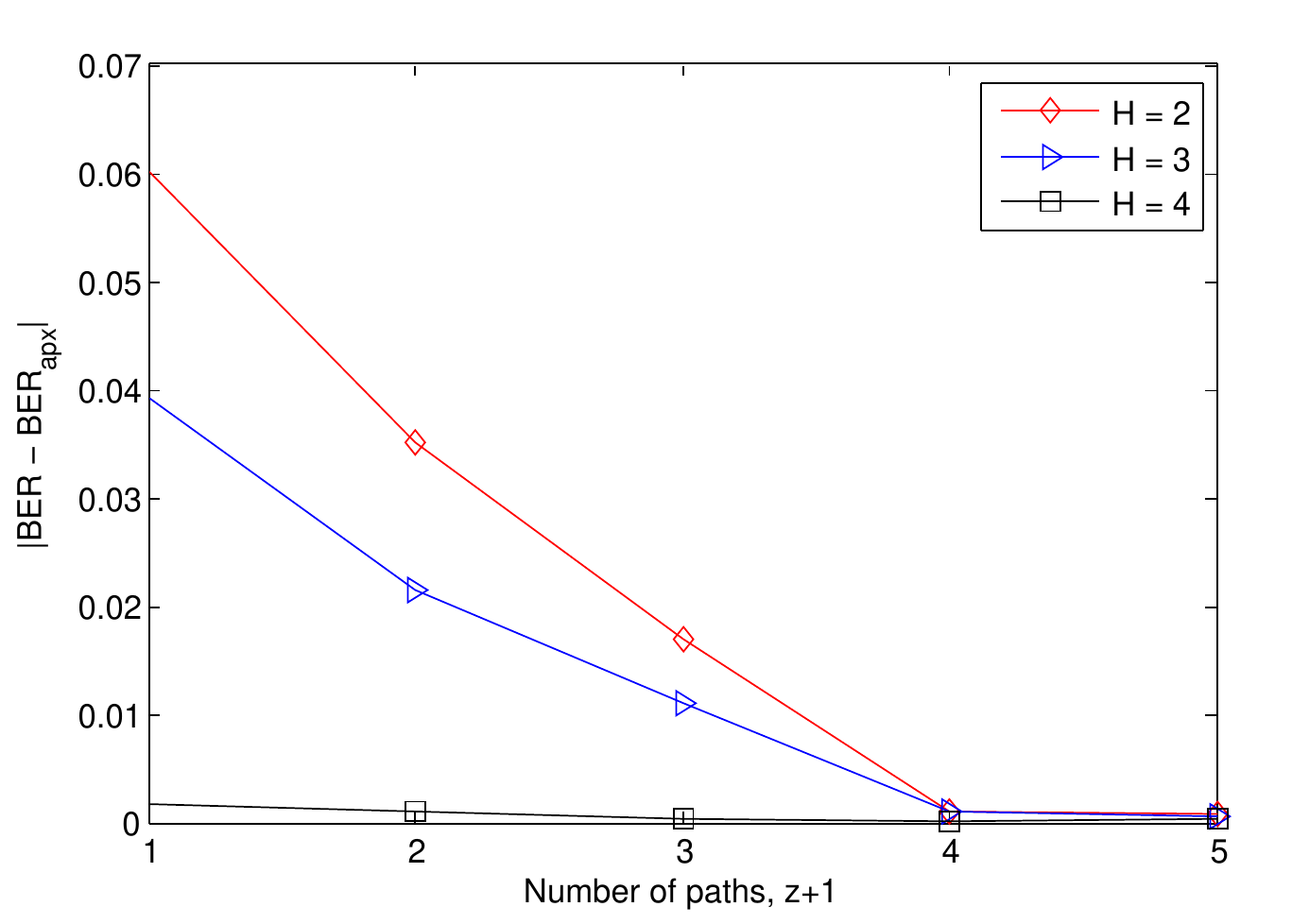}
\caption{\small $|BER - BER_{apx}|$ for different PTC ($\mathit{H}$ = 2, 3, and 4) and SNR = 7 dB.
}\label{fig:diff}
\end{figure}

In Fig. \ref{fig:upr_b}, we fix $H=3$ and compare the simulated BER with its approximations using different number of paths used in the approximation, i.e., $z+1$. It is clear from the figure that $z=3$ approximates BER well for $H=3$. In Fig. \ref{fig:diff}, we fix $SNR=7$ dB and compare absolute error of the BER approximation for different values of $H$. It is again clear that $z=3$, i.e., using only $z+1=4$ paths of the convolutional code's trellis, provides good BER approximation for different sizes of PTC. This is due to the fact that the first few error events with a few errors in the trellis dominate all the remaining transitions. This result confirms what we have mentioned earlier regarding the domination of a small number of paths in the trellis over all the remaining paths that may evolve due to incoming bits. 
The approximate BER we provide in the section could be used as a QoS metric in an adaptive scenario where certain parameters of the system such as $H$ could be adapted in an online manner using this approximation.
In Table~\ref{timetb}, we present results on the performance of exhaustive search\footnote[1]{Using this technique, decisions on the transmitted code matrices are made by the permutation trellis decoder which decides in favor of the transmitted code matrix, i.e., among $2^m$ possible code matrices, which has the minimum Hamming distance with respect to the received one. In order to determine all the decisions for every possible code matrix, a brute force method is used to compare each possible received code matrix to
every code matrix. This method needs roughly $2^{H^2 + m}$ comparisons~\cite{ragg}.} compared to the proposed approximation approach in this paper in terms of the time needed to compute BER for different PTC. \begin{table}[t]
\caption{\small A comparison of the time needed to compute BER between exhaustive search~\cite{ragg} and approximation approach for different PTC ($\mathit{H}$ = 2,3, and 4)}
\label{timetb}
\begin{center}
    \begin{tabular}{| c | c | c |}
    \hline
PTC  & Exhaustive Search Approach & Approximate Approach \\ \hline
$H = 2$ & $0.33 s$ & $0.31 s$\\ \hline
$H = 3$ & $42.21 s$ & $7.24 s$\\ \hline
$H = 4$ & $26274.99 s$ & $245.32 s$\\ \hline
\end{tabular}
\end{center}
\end{table} As shown in Table~\ref{timetb}, it takes much shorter time to compute the BER approximately as compared to its exact calculation.
\subsection{Throughput Analysis of $H$-FSK}
For comparison purposes, we first consider a competing scenario where an SU does not employ any coding technique in its spectrum interweaved transmission strategy which, in turn, is assumed to be based on $M$-FSK modulated signals. In this scenario, an SU makes use of the CR channel sensing feature in order to determine if there is an active PU in the network or not. Depending on the channel sensing measurements of the licensed spectrum, a decision is made on whether the SU should vacate the band or not. In the case where the SU detects a white space, it adapts the transmission parameters to utilize the full spectrum available and achieve the maximum throughput. On the other hand, when a PU is detected an SU vacates the band and adjusts its transmission parameters accordingly by employing a modified FSK modulation scheme supporting the transmission of $M'$ symbols where $M'$ $<$ $M$.
\begin{figure}[t]
\centering
\includegraphics[width=0.6\textwidth,height=!]{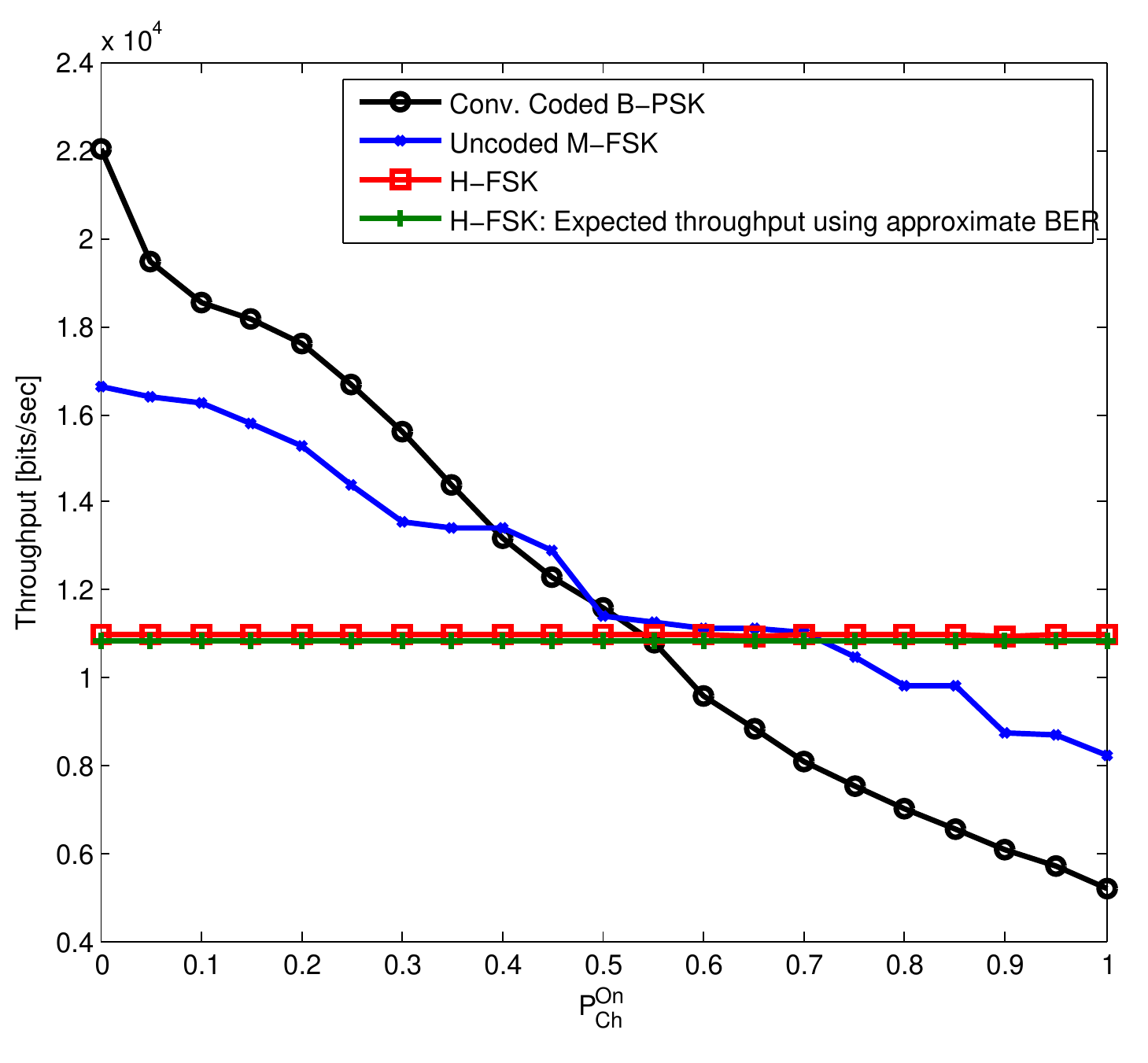}
\caption{\small Throughput Comparison of $H$-FSK and uncoded opportunistic $M$-FSK as well as convolutionally coded BPSK schemes for SNR = 4 dB, packet size = 256 bits, and $R_p$ = 100 packets/sec.
}\label{fig:goodput}
\end{figure}
In order to gain more insight regarding the throughput performance comparison of the PTC-based multi-level FSK with the uncoded-opportunistic (adaptive) $M$-FSK system, we assume that packet transmissions take place over an observation window of length $T$ seconds divided into $x$ time slots. For each time slot, the steady state probability of a channel being occupied by an active PU, given in (\ref{sson}), is evaluated. Knowing each interval's time and the time required for transmitting a single packet, the incorrectly received packets can be determined. As a result, the unlicensed network's throughput is computed and analyzed. We further assume that the available spectrum consists of four frequency bands. With the use of the PTC-based multi-level FSK approach, an SU does not need to sense the channel or adapt its wireless links. In the uncoded opportunistic system, an SU senses the licensed frequency band at the beginning of a time slot. Assuming zero-cost perfect detection of a licensed user's activity, an SU decides on which $M$-FSK modulated signals are to be transmitted. In the case where an SU detects a spectrum hole, it utilizes the full available spectrum (4 bands) using a 4-FSK modulation scheme. When a PU is detected, an SU vacates the licensed band and utilizes the remaining spectrum (2 frequency bands out of the 4 bands assumed to comprise the spectrum) using BFSK modulated signals. Different from \cite{puu}, the licensed channel can change its state at any instant in a time slot. In Fig. \ref{fig:goodput}, we present the average throughput of $H$-FSK and that of an uncoded opportunistic $M$-FSK system as a function of  $P_{ch}^{On}$  for $H = M = 4$ where the packet size $L=256$ bits, $R_p$ = 100 packets/sec, and SNR is 4 dB. Fig. \ref{fig:goodput} illustrates that the average throughput in $H$-FSK is constant regardless of $P_{ch}^{On}$ values while the average throughput of the uncoded opportunistic $M$-FSK decreases as $P_{ch}^{On}$ increases and it matches the theoretical approximation given in \eqref{eq:expth}. As shown, there exists a probability $p_1^* \approx 0.7$ such that for $P_{ch}^{On}$ $\geq$ $p_1^*$ the $H$-FSK approach outperforms the uncoded opportunistic scheme in terms of throughput performance besides BER. Thus, the use of the proposed $H$-FSK communication scheme for values of $P_{ch}^{On}$ less than $p_1^*$ is not beneficial because higher throughput values are obtained using the uncoded opportunistic scheme. In other words, the $H$-FSK system is effective for situations where the PU activity is relatively high. In practice, the throughput evaluated for the uncoded $M$-FSK system is expected to be lower than what is shown in Fig. \ref{fig:goodput} due to channel sensing intervals. It should be noted that the proposed $H$-FSK communication scheme does not need the channel sensing mechanism that might not be available for some applications.

Along these lines, we compare the average throughput of the proposed $H$-FSK scheme to a rate-$\frac{1}{2}$ convolutionally coded BPSK modulated OFDM system. We assume that the latter transmits a BPSK modulated signal in parallel over the available frequency bands in a given time window similar to what is considered in the previous comparison setting. For fair comparison, neither scheme carries out sensing, i.e., they transmit at all times. It is worth noting that the throughput performance of this BPSK OFDM system is investigated assuming the same transmission bandwidth, power, and time frame as with the $H$-FSK system. As shown in Fig. \ref{fig:goodput}, the average throughput of the convolutionally coded BPSK system also decreases as $P_{ch}^{On}$ increases. Similar to the previous throughput comparison, there exists a probability $p_2^* \approx 0.55$ such that for $P_{ch}^{On}$ $\geq$ $p_2^*$ the $H$-FSK system outperforms the convolutionally coded BPSK scheme in terms of average SU's throughput. Therefore, the use of the proposed $H$-FSK communication scheme for values of $P_{ch}^{On}$ less than $p_2^*$ is not beneficial because higher throughput values would be obtained using the other coded scheme.
\subsection{SU Link's Resiliency in the Presence of Multiple PUs}
In this section, we further carry out our performance analysis and examine the resiliency of the network (SU's link) to interference provided by the proposed PTC based framework in the presence of multiple PUs when i) PUs \textit{always} transmit and ii) PUs are \textit{dynamic} and their activities vary dynamically. It is worth noting here that the PUs are located $10 m$ away from the SU receiver in order to analyze the quality of the SU link under a high interference scenario.
\begin{figure}[t]
\centering
\includegraphics[width=0.6\textwidth,height=!]{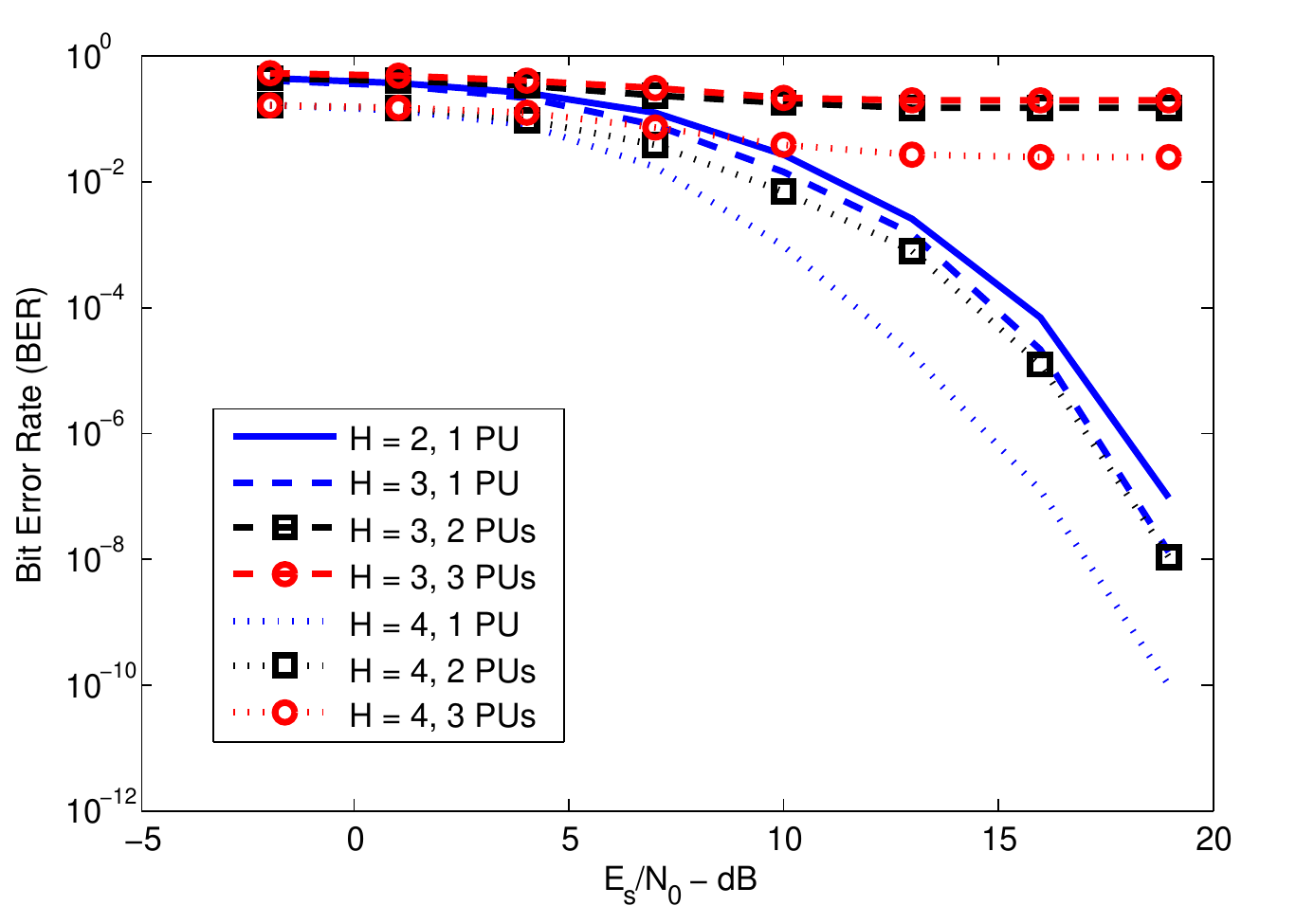}
\caption{\small Approximate BER performance of different PTCs with respect to PUs that are \textit{always} ON and transmitting.
}\label{fig:prev_res}
\end{figure}
\begin{figure}[t]
\centering
\includegraphics[width=0.6\textwidth,height=!]{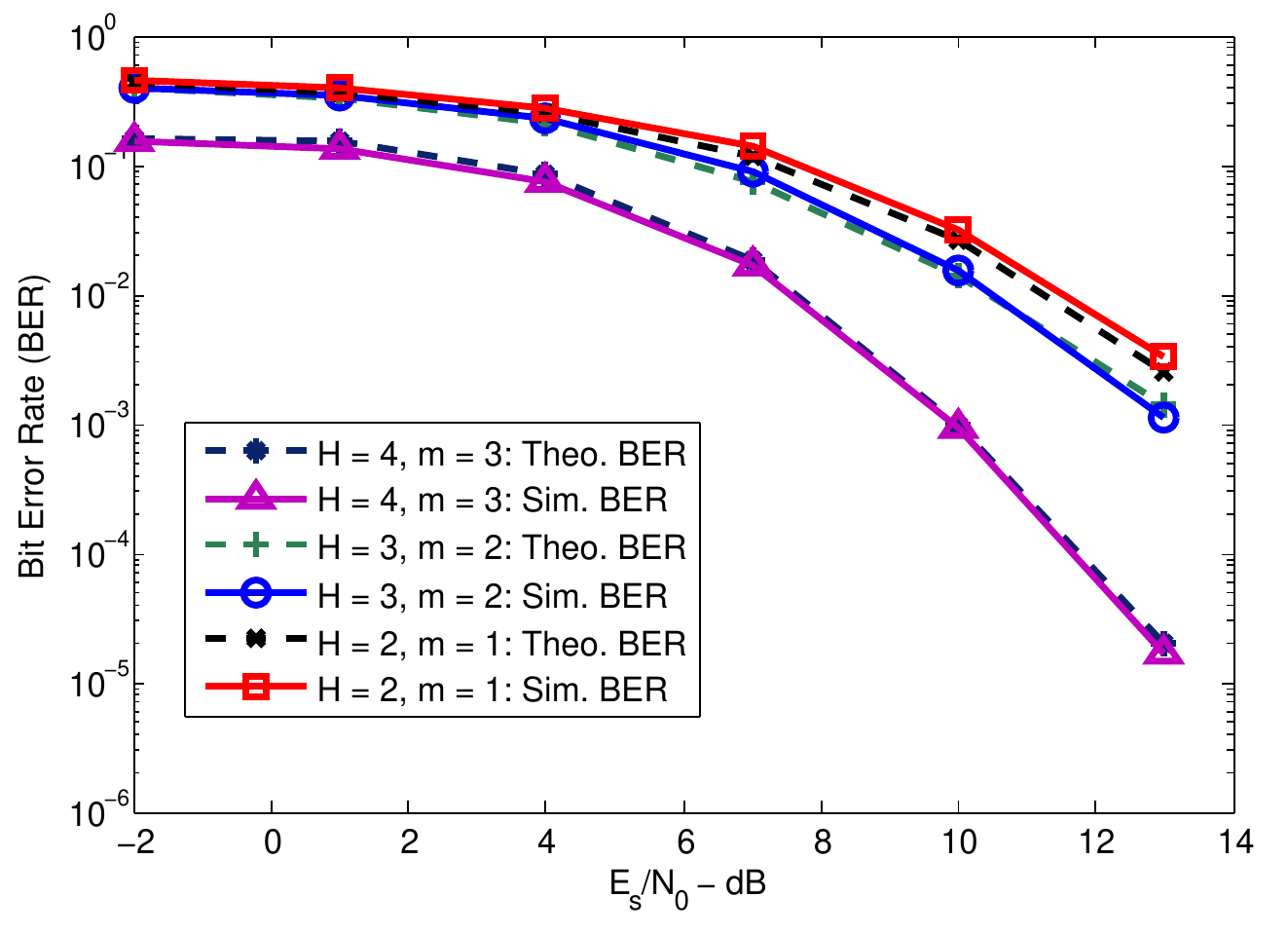}
\caption{\small Approximate BER  versus Simulated BER for a PTC with $H = 2,$ $3,$ $\mbox {and } 4$ and $P_{ch}^{ON} = 0.35$.
}\label{fig:pu_dyn}
\end{figure}
\begin{figure}[t]
\centering
\includegraphics[width=0.6\textwidth,height=!]{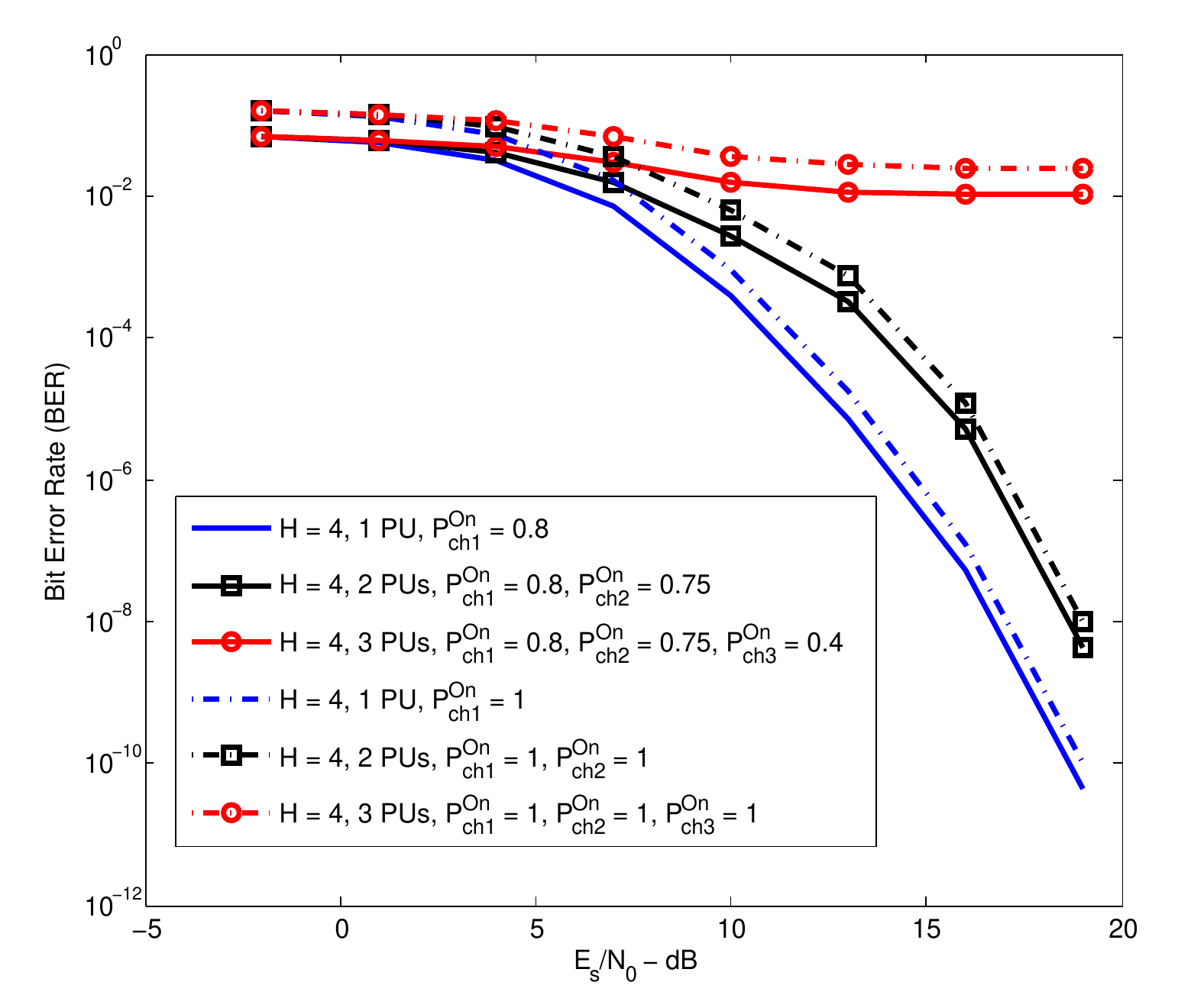}
\caption{\small Approximate BER performance of a PTC ($H$ = 4) with respect to dynamically varying PUs activities and PUs that are \textit{always} ON.
}\label{fig:mult_ber}
\end{figure}

\begin{figure}[t]
\centering
\includegraphics[width=.6\textwidth,height=!]{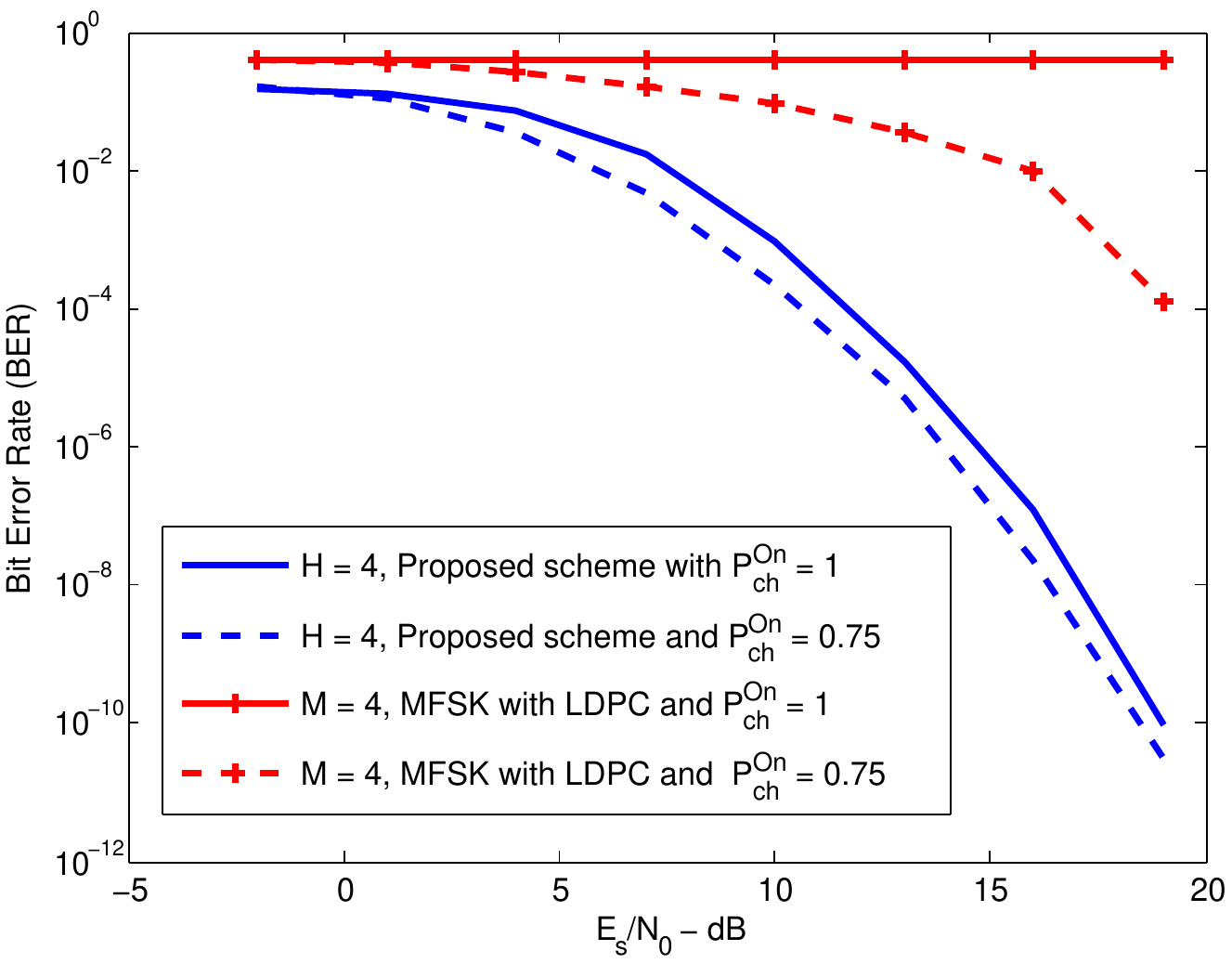}
\caption{\small A BER comparison between the proposed PTC scheme with $H = 4$ and a $1/2$ rate ($64800$,$32400$) LDPC encoding with a $4$-FSK modulation system with respect to static ($P_{ch}^{On}$ = 1) and dynamic ($P_{ch}^{On}$ = 0.75) activities of a PU.
}\label{fig:newplot}
\end{figure}
\subsubsection{\textit{Static} Channel Occupancy}
First, we consider multiple PUs that are \textit{always} active in the network during SU communication. In~Fig. \ref{fig:prev_res}, we plot the approximate BER performance of the SU link for different PTCs in the presence of multiple PUs. The results obtained in the presence of multiple PUs that are \textit{always} ON are similar to those presented in~\cite{rf11,rf10}. These results show that the use of larger PTC (higher value of $H$) adds more robustness to SU communications, i.e., BER decreases as the value of $H$ increases. It is also shown in these results that the approximate BER performance of an SU link degrades as the number of PUs joining the network increases. Of course, this is achieved at the expense of larger overhead when larger values of $H$ are used.
\subsubsection{\textit{Dynamic} Channel Occupancy}
The case of \textit{static} PUs that \textit{always} transmit is more of a pessimistic scenario as presented above. In this paper, we extend the work in~\cite{rf11,rf10} and consider a more practical scenario to model the intermittent \textit{dynamic} activities of PUs in the network. Using a 2-state Markov chain to model channels' occupancy by PUs, we provide a more accurate characterization of system performance and verify the approximate BER in the presence of a dynamically transmitting PU with simulated BER for a given PTC and known parameters of the channel occupancy model. In~Fig.~\ref{fig:pu_dyn}, we present the approximate BER performance of the proposed framework as a function of $\frac{E_s}{N_0}$ for different PTC schemes, i.e., $H = 2, 3, \mbox{ and }4$, and $P_{ch}^{ON} = 0.35$. Both analytical and simulation results are presented which match each other quite well.

In addition, in~Fig.~\ref{fig:mult_ber}, we further examine the approximate BER performance of SU communications for a given PTC, i.e., $H = 4$, in the presence of one, two, and three PUs that i) are \textit{static} in nature and \textit{always} transmit and ii) \textit{dynamic} in nature. The results obtained in Fig.~\ref{fig:mult_ber} clearly show that there exists a gap between the performance of the systems in both cases. Our analysis based on a more practical occupancy model is able to predict SU communication performance more accurately and which is better than the pessimistic case of PUs being \textit{always} ON. For example, the approximate BER value in the presence of one \textit{dynamic} PU is lower than when a PU is \textit{always} transmitting. This is intuitive because only a fraction of the transmitted information packets are affected by interference from the \textit{dynamic} PU when it is transmitting compared to the \textit{static} PU case where all the transmitted information packets experience permanent interference. In fact, this is also observed for the case of 2 and 3 PUs being active in the network.

Moreover, we compare the BER performance of the proposed communication system when $H = 4$ with another system that is based on a $1/2$ rate ($64800$,$32400$) LDPC encoding given in Matlab toolbox with a $4$-FSK modulation system. Fig. \ref{fig:newplot} illustrates how the proposed communication system outperforms the aforementioned system for both static and dynamic activities of a PU.

\section{Conclusion}
\label{sec:conc}
In this paper, we employed a PTC based framework to mitigate the impact of PUs modeled using a practical \textit{dynamic} channel occupancy model in CRNs. We, further, computed the SU link's BER approximately which was shown to be quite accurate. This approximation allows one to use BER as a QoS metric to determine the link quality of an SU link for applications such as link adaptation. Furthermore, in order to assess the effectiveness of the proposed PTC based multi-level FSK communication scheme, we compared the performance to that of an uncoded opportunistic $M$-FSK system, a coded $M$-FSK system, and a coded BPSK modulated system deploying parallel transmissions. Based on the comparisons described in Section~\ref{sec:NRD}, we showed that the proposed scheme outperforms the latter two under relatively heavy PU interference. We also presented results that exhibit the resiliency of an SU link to interference for PTCs in the presence of multiple \textit{dynamic} PUs activities.

In this work, we evaluated the performance of PTC codes for a single SU link. Optimal code assignments in a multiple SU scenario will be considered as a future research direction. We also plan to consider 1) more realistic channel models that account for shadowing and fading effects; and 2) more effective decoding techniques, e.g., soft-decision decoding. Based on the promising results we have obtained in a CRN, we intend to generalize the proposed scheme to tackle wireless networks subject to jamming attacks and malicious nodes' activities.


\appendix
\section*{Proof of Proposition 1}
Let $\yp(\ccc_1)$ be the transmitted codeword and $\R^{(i)}$ be an arbitrary received vector. Assume that $\yp(\ccc_2)$ is another possible codeword. If there exists a received vector $\R^{(j')}$ such that $d_H(\R^{(j')},\yp(\ccc_2))=d_H(\R^{(i)},\yp(\ccc_1))$ and $P(\R^{(j')}|\yp(\ccc_2))=P(\R^{(i)}|\yp(\ccc_1))$, then the upper-bound given in \eqref{berr2} is the same no matter what the transmitted codeword is ($\yp(\ccc_2)$ or $\yp(\ccc_1)$).
Let $\sss_1=(s_{1,\ell H^2},s_{1,\ell H^2+1},\ldots, s_{1,(\ell+1) H^2-1})$ be a portion of $\yp(\ccc_1)$ generated from the corresponding $(\ell+1)$th output branch of the convolutional code and $\sss_2$ be that of $\yp(\ccc_2)$ for the same bit positions. Since the number of non-zero elements in $\sss_1$ and $\sss_2$ is $H$, we can assume that they occupy the following bit indices $o_{1,0},o_{1,1},\ldots, o_{1,H-1}$ in $\sss_1$ and $o_{2,0},o_{2,1},\ldots, o_{2,H-1}$ in $\sss_2$, respectively. Let $\rrr_1=(r_{1,\ell H^2},r_{1,\ell H^2+1},\ldots, r_{1,(\ell+1) H^2-1})$ be the corresponding received vector of $\sss_1$. Note that $\rrr_1$ here is in the form of an arbitrary received code matrix for each output branch of the convolutional code and is mapped to a permutation trellis code matrix. Then, $\R^{(j')}$ can be constructed from $\R^{(i)}$ as provided below.\\
Define a one-to-one mapping $\pi_\ell$ from indices $\{\ell H^2,\ell H^2+1,\ldots, (\ell+1) H^2-1\}$ to itself as
\begin{eqnarray}
\pi_\ell(q)=\left\{\begin{array}{cl}
o_{2,i},&\mbox{ if } q= o_{1,i} \mbox { for } 0\le i\le H-1,~\\
o_{1,i},&\mbox{ if } q= o_{2,i} \mbox { for } 0\le i\le H-1,~\\
q,&\mbox{ otherwise.}
\end{array}\right.\nonumber
\end{eqnarray}
It is easy to see that the above mapping rule switches $\sss_1$ to $\sss_2$ and vice versa. We can write $\pi(\yp(\ccc_1))=\yp(\ccc_2)$ and $\pi(\yp(\ccc_2))=\yp(\ccc_1)$. At the same time, the mapping rule also switches $r_{1,o_{1,i}}$ with $r_{1,o_{2,i}}$ for $0\le i\le H-1$ when applied to $\rrr_1$. Then, the combination of $\ell$ mappings in $\pi$ with respect to $\R^i$, results in
$$\R^{(j')}=\pi(\R^{(i)})=(r_{1,\pi_0(0)}, r_{1,\pi_0(1)},\ldots,r_{1,\pi_0(H^2-1)},r_{1,\pi_1(H^2)},\ldots,r_{1,\pi_{L+m-1}((L+m-1)H^2-1)}).$$ Accordingly, $$d_H(\R^{(i)},\yp(\ccc_1))=d_H(\pi(\R^{(i)})),\pi(\yp(\ccc_1)))=d_H(\R^{(j')}),\yp(\ccc_2))$$
for the one-to-one mapping conserves the Hamming distance of the convolutional code. In the case where noise for each received bit is independent, we have
$$P(\R^{(i)}|\yp(\ccc_1))=P(\pi(\R^{(i)})|\pi(\yp(\ccc_1)))=P(\R^{(j')}|\yp(\ccc_2))$$
and since noise of the received bits in a subchannel occupied by the PU is not independent, $P(\R^{(i)}|\yp(\ccc_1))$ is not always equal to $P(\R^{(j')}|\yp(\ccc_2))$. However, if we assume that the noise caused by the PU in a subchannel will affect all bits in this channel that are generated by one or more trellis branches, then $$P(\R^{(i)}|\yp(\ccc_1))=P(\pi(\R^{(i)})|\pi(\yp(\ccc_1)))=P(\R^{(j')}|\yp(\ccc_2)).$$
We can reason this from the fact that bit-switching by $\pi$ on $\R^{(i)}$ does not move any bit out of the row it is located in. 

\section*{Acknowledgment}
This material is based on research sponsored in part by CASE: The Center for Advanced Systems and Engineering, an NYSTAR center for advanced technology at Syracuse University.
Part of Han's work is supported by National Science Council, Taiwan, R.O.C., under grants NSC 101-2221-E-011 -069 -MY3 and NSC 99-2221-E-011 -158 -MY3.

\bibliography{references}
\bibliographystyle{IEEEtran}

\end{document}